\documentclass{appolb}
\usepackage{epsfig}

\begin{document}
\title{Gravitational field energy contribution to the neutron star mass
 \thanks{Partially supported by the Polish Ministry of Science and Informatization, grant 1P03D00528}
}
\author{{M.Dyrda$^a$, B.Kinasiewicz$^a$, M.Kutschera$^{a,b}$, A. Szmagli\'nski$^c$}
\address{~$^a$ The M. Smoluchowski Institute of Physics, Jagellonian University, Reymonta 4, 30-059 Krak\'ow, Poland}
\address{~$^b$ The H. Niewodnicza\'nski Institute of Nuclear Physics, Radzikowskiego 142, 31-342 Krak\'ow, Poland}
\address{~$^c$ Institute of Physics, Technical University, Podchor\c{a}\.zych 1, 30-084 Krak\'ow, Poland}}

\maketitle
\begin{abstract}
Neutron stars are discussed as laboratories of physics of strong gravitational fields. The mass of a neutron star is split 
into  matter energy and gravitational field energy contributions.
The energy of the gravitational field of neutron stars is calculated with three different approaches which give the same result.
It is found that up to one half of the gravitational 
mass of maximum mass neutron stars is comprised by the 
gravitational field energy. Results are shown for a number of realistic equations of state of neutron star matter.  
\end{abstract}
\PACS{04.40.Dg, 97.60.Jd}

\section{Introduction}
Gravity makes all selfgravitating fluid bodies spherical. Thus the spherical symmetry of non-rotating
neutron stars, which are real objects, is their intrinsic property, not an idealization. 
Neutron stars are relativistic objects, with mass of order $1M_{\odot}$ enclosed within 10-20 km.  Gravity of such
a compact mass distribution is strong enough to require relativistic description. In the following we focus on
gravitational field contribution to the neutron star masses in General Relativity.

The problem of the energy distribution of gravitational field remains an unsettled issue in general relativity
\cite{Cooperstock} mainly because of the lack of proper energy-momentum tensor for gravitational field. 
However, for static, spherically symmetric space-time no controversy exists  and relevant
matter and gravity energies can be unambiguousely defined \cite{MTW, Weinberg}. Our main concern here is to study if
neutron stars could shed some new light on the problem of gravitational field energy and its localization.   

To calculate both matter and gravitational field  contributions to the neutron star mass we solve the Einstein's 
field equations for
the static, spherically symmetric metric and a given equation of state of dense matter.
The metric is
\begin{equation}
{ds}^2 = e^{\nu }c^2{dt}^2 - e^{\lambda }{dr}^2- r^2\left({d\theta}^2 + {\sin{\theta}^2}{d\phi}^2 \right),
\label{metrics} 
\end{equation}
where $\nu$, $\lambda$ are functions of the radial coordinate $r$.

Dense matter in neutron stars is thought to be in a liquid state, hence it is described by the perfect fluid 
energy-momentum tensor,
\[T^{\mu \nu}=(\rho c^2 + P)u^{\mu}u^{\nu}-Pg^{\mu \nu},\] where $\rho c^2$ is the energy denisty, $P$ is the 
pressure and $u^{\mu}$ is the four velocity. For the neutron star matter both energy density and pressure are 
uniquely determined by the baryon number
density, $\rho=\rho(n_B), P=P(n_B)$, hence the equation of state is of barotropic form, $P=P(\rho)$. 
In this paper we employ a number of realistic equations of state, described in some details in Sect. \ref{EOSNM}.

Using the Einstein's equations one finds that functions $\lambda$ and $\nu$ 
satisfy ordinary differential equations

\begin{eqletters}
\label{myeq}
\begin{eqnarray}
\lambda'(r)= {1 \over r}\left(1 - e^{\lambda } + {8\,\pi\,G\,e^{\lambda }\,r^2\,\rho(r) \over c^2}\right)
\label{me1}
\\
\nu'(r)={1 \over r}\left(-1 + e^{\lambda } + {8\,\pi\,G\,e^{\lambda }\,r^2\,P(r) \over c^4}\right)
\label{me2}
\\
P'(r)=-{1 \over 2}(P(r) + \rho c^2)\nu'.
\label{me3}
\end{eqnarray}
\end{eqletters}

The metric (1) has to be continuous in the whole space. For this to happen, at the stellar boundary, $r=R$, 
the internal metric has to go over into the external Schwarzschild metric. There is a vacuum
outside the star and the matter energy-momentum tensor vanishes. The metric functions at the boundary should 
satisfy the condition $e^{\nu(R)}=e^{-\lambda(R)}=1-2GM/Rc^2$, where $M$ is the mass of the star.

From Eqs.(2) one derives the Tolman-Oppenheimer-Volkoff (TOV) equation that describes  nonrotating compact objects 
in the hydrostatic equilibrium:
\begin{eqnarray}
\lefteqn{ {dP(r)\over dr}= - {GM(r)\rho(r)\over r^2}\times } \nonumber \\ & &
\left(1+ {P(r)\over c^2\rho(r)}\right)\left(1+{4\pi r^3P(r)\over c^2M(r)}\right)\left(1-{2GM(r)\over c^2r}\right)^{-1}
\end{eqnarray}
Here we denote by $M(r)$ the integral $M(r)=\int_0^r\rho(r')d^3r'$. Usually $M(r)$ is referred to as "the mass within 
the radius $r$". This name is rather misleading, as it suggests that only matter energy density is responsible for
building the neutron star mass $M=M(R)$. It is our prime goal here to calculate the gravitational energy contribution 
to $M$. In the following we shall be using the differential form of the $M(r)$ definition,
\begin{equation}
	{dM(r)\over dr}= 4\pi r^2\rho(r).
\label{eqM}
\end{equation}
Choosing an appropriate equation of state of dense matter
\begin{equation}
	P = P(\rho),
\end{equation}
and the boundary conditions we  solve this set of equations for different values of central density $\rho_c$ obtaining the whole family of configurations of compact objects in hydrostatic equilibrium.
\begin{equation}
	\rho(r=0)=\rho_c \ \mathrm{and} \ P(r=R)=0,
\end{equation}

\begin{figure}[!hbt]
	\begin{center}
		\includegraphics[height=5.5cm]{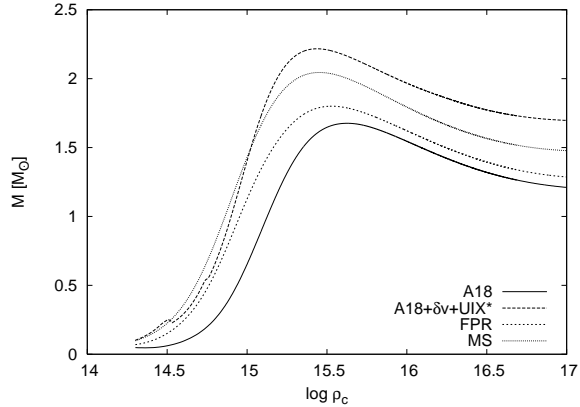}
	\end{center}
	\caption{Neutron star masses \textit{versus} central density for four equations of state.}
	\label{fig:ns_konf}
\end{figure}
We report here calculations for four equations of state, which we denote as A18, A18+$\delta$v+UIX*, FPR and MS (see Sect. \ref{EOSNM}). These four EOS have been chosen from about twenty EOS we studied solely to make the plots clear.
The results of our calculations are shown in  Fig. \ref{fig:ns_konf}.

Numerical values of the neutron stars masses for these equations of states are given in the Table \ref{tab:masses}. 
We show parameters corresponding to the standard neutron star mass $M=1.44M_{\odot}$, and to the maximum mass neutron
star, which is different for every equation of state.  
\begin{table}[t]
\begin{center}
\begin{tabular}{l c c }
              
Equation of state   &  Central density [$10^{15}\,g/cm^3$] & Mass [$M_\odot$]   \\  \hline 
A18 & 2.54 & $1.44$   \\
A18 & 4.23&$1.68$  \\ \hline

A18+$\delta$v+UIX* & 1.02 & $1.44$  \\
A18+$\delta$v+UIX* & 2.74 & $2.22$ \\ \hline
FPR & 1.36 & $1.44$ \\ 
FPR & 3.38 & $1.80$ \\ \hline 
MS & 1.01 & $1.44$ \\ 
MS & 2.84 & $2.04$ \\  \hline
\end{tabular}
\caption{Computed neutron star masses. For each equation of state the first line shows the value of the central 
density of $1.44\ M_{\odot}$ neutron star, and in the second row the central density of the maximum mass 
neutron star and the value of this mass are given.
\label{tab:masses}}
\end{center}
\end{table}

\section{Gravitational field energy of the compact objects}
Boundary conditions at the surface of the neutron star, $r=R$, ensure that the external metric is just the Schwarzschild
metric with the mass $M$. There is no more matter contributions to $M$ at $r>R$ since the matter energy density is zero
outside the star. Hence the same mass $M$ survives in the metric functions in the asymptotic region, $r \to \infty$, and
\ $M$ is the
total gravitational mass of the neutron star (sometimes called the ADM mass).

In General Relativity the mass of the star can be split into two parts: the matter energy and the gravitational field 
energy \cite{Tolman}, $Mc^2=E_{Matter}+E_{Field}$. Formally we can write
\begin{equation}
Mc^2=4\pi  \int_0^R(T_0^0+t_0^0)e^{{\nu/2}} e^{{\lambda/2}} r^2 dr,
\end{equation}
where $T_0^0$ is the energy density component of the matter energy-momentum tensor, and $t_0^0$ is the energy density of the 
gravitational field pseudotensor.

The matter energy can be easily calculated for any given neutron star, as all three functions, $\rho(r), \lambda(r), \nu(r)$
entering the integral, 
\begin{equation}
E_{Matter}=4\pi c^2 \int_0^R \rho\, e^{{\nu/2}} e^{{\lambda/2}} r^2 dr,
\label{E_Matter}
\end{equation}
are determined by solving Eqs.(2). One should notice that not only the integral (\ref{E_Matter}) is uniquely determined, 
but also matter energy density is completely known, $\rho_{Matter}=\rho e^{\nu/2}$. This energy density includes contributions
from nuclear and electromagnetic interactions (through $\rho$), as well as the gravitational interaction (through $e^{\nu/2}$). 
It is important to stress that, in the Newtonian limit, the generally relativistic matter energy $E_{Matter}$ includes twice
 the (negative) potential energy contribution, as shown by Lynden-Bell and Katz \cite{LBK}.  

\begin{figure}[hbt]
	\begin{center}
		\includegraphics[height=6.5cm]{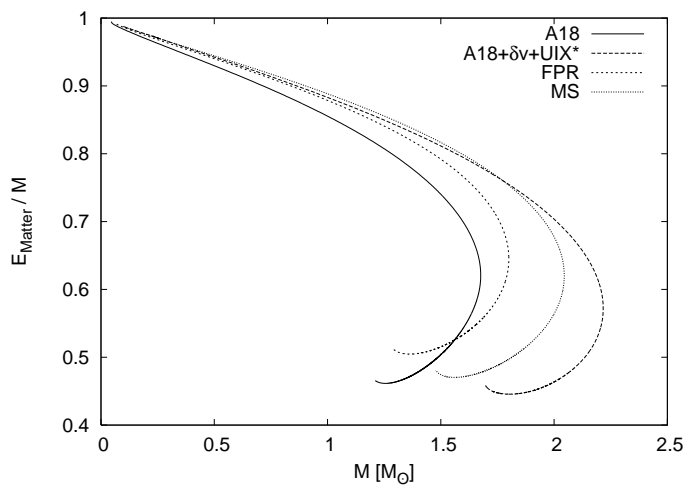}
	\end{center}
	\caption{Matter energy contribution to the neutron star mass as a function of the total gravitational mass.}
	\label{fig:ns_matter_energy}
\end{figure}

Using this definition (\ref{E_Matter}) we can compute the matter energy for all equations of state we study. The results of 
our calculations are shown in  the Fig. \ref{fig:ns_matter_energy} as functions of the total gravitational mass.  

One should perhaps stress once again, that for a given equation of state and a chosen central density, the matter part of the
total mass integral (7) is fully determined. For spherically symmetric metric there does not exist any ambiguity as far as the
integral (8) is concerned, despite the pseudotensorial nature of the gravitational field energy density $t_0^0$. The equations
(2) are fully immune to the particular choice of the gravitational field energy-momentum pseudotensor.

Having the matter energy $E_{Matter}$ we can easily find the gravitational field energy contribution to the neutron star mass,
\begin{equation}
E_{Field} = Mc^2 - E_{Matter} = 4\pi c^2\int_0^R\left(e^{-{\lambda/2}}-e^{{\nu/2}}\right)\rho e^{{\lambda/2}} r^2 dr
\label{eg1}
\end{equation}
where, formally, we can refer to the function under the integral as the "gravitational field energy density",
\begin{equation}
\rho_{Field}(r)=(e^{-{\lambda\over 2}}-e^{{\nu\over 2}})\rho(r), 
\end{equation}
in units of $g/cm^3$. Unfortunately, in contrast to the gravitational field energy contribution to the neutron star 
mass, $E_{Field}$, which is fully determined from Eq.(\ref{eg1}), this function is not unique. The reason is that $t_0^0$
is not unique, only the integral in Eq.(\ref{eg1}) has a well defined meaning.

\begin{figure}[hbt]
	\begin{center}
	
\includegraphics[height=4.3cm]{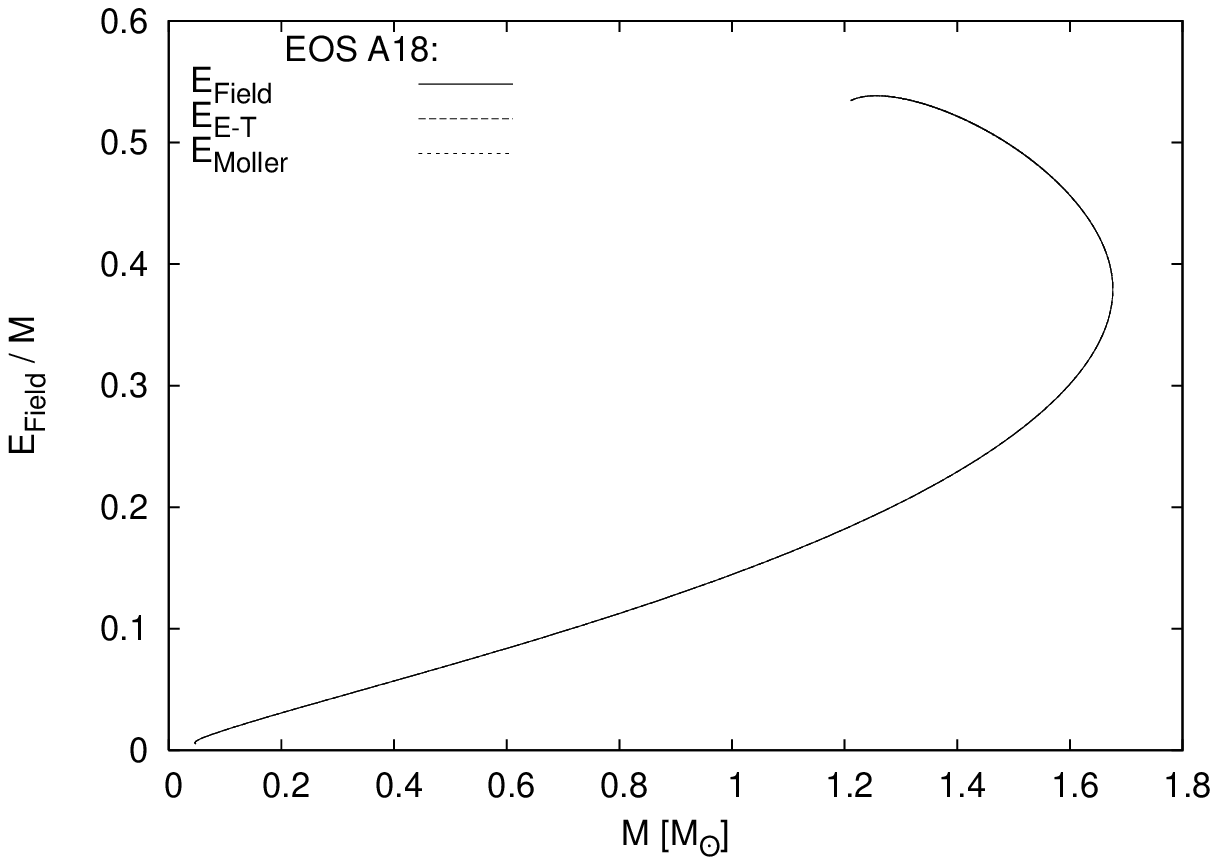}\includegraphics[height=4.3cm]{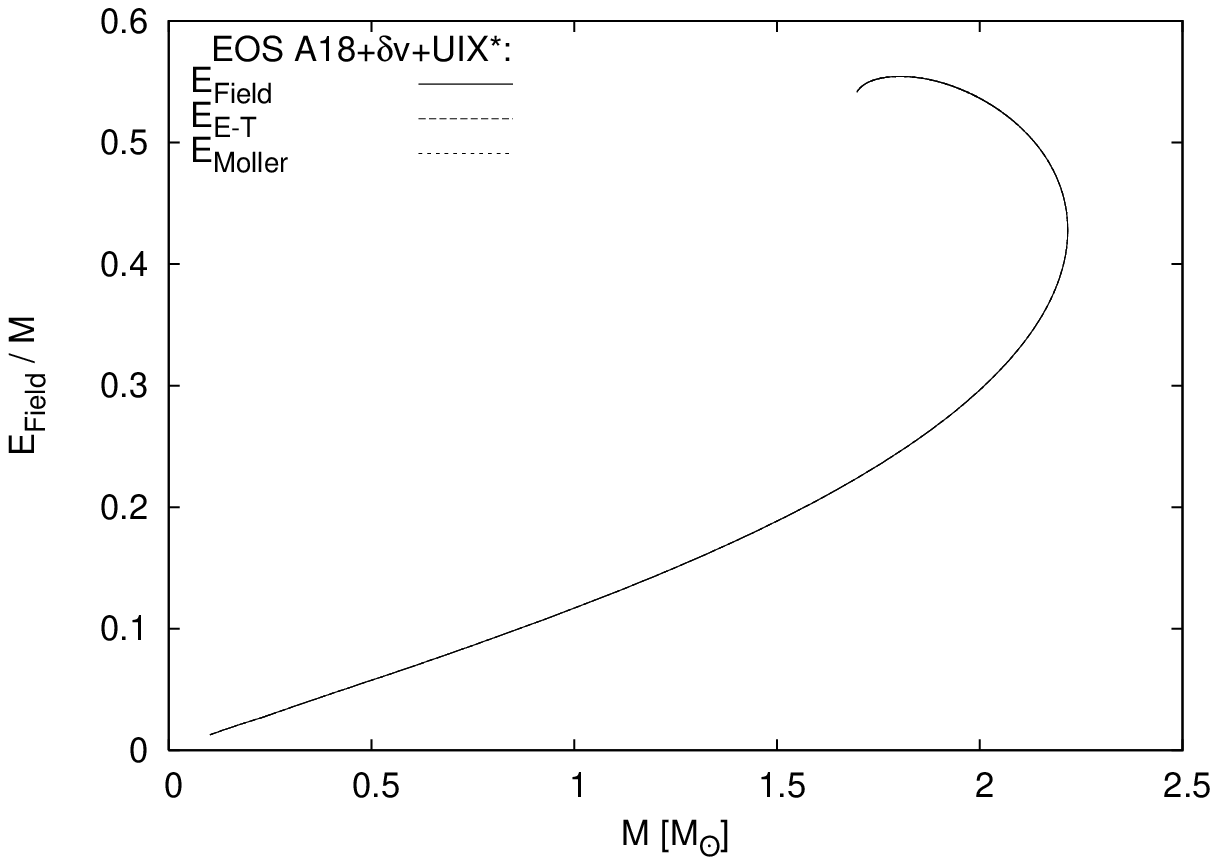}
	
\includegraphics[height=4.3cm]{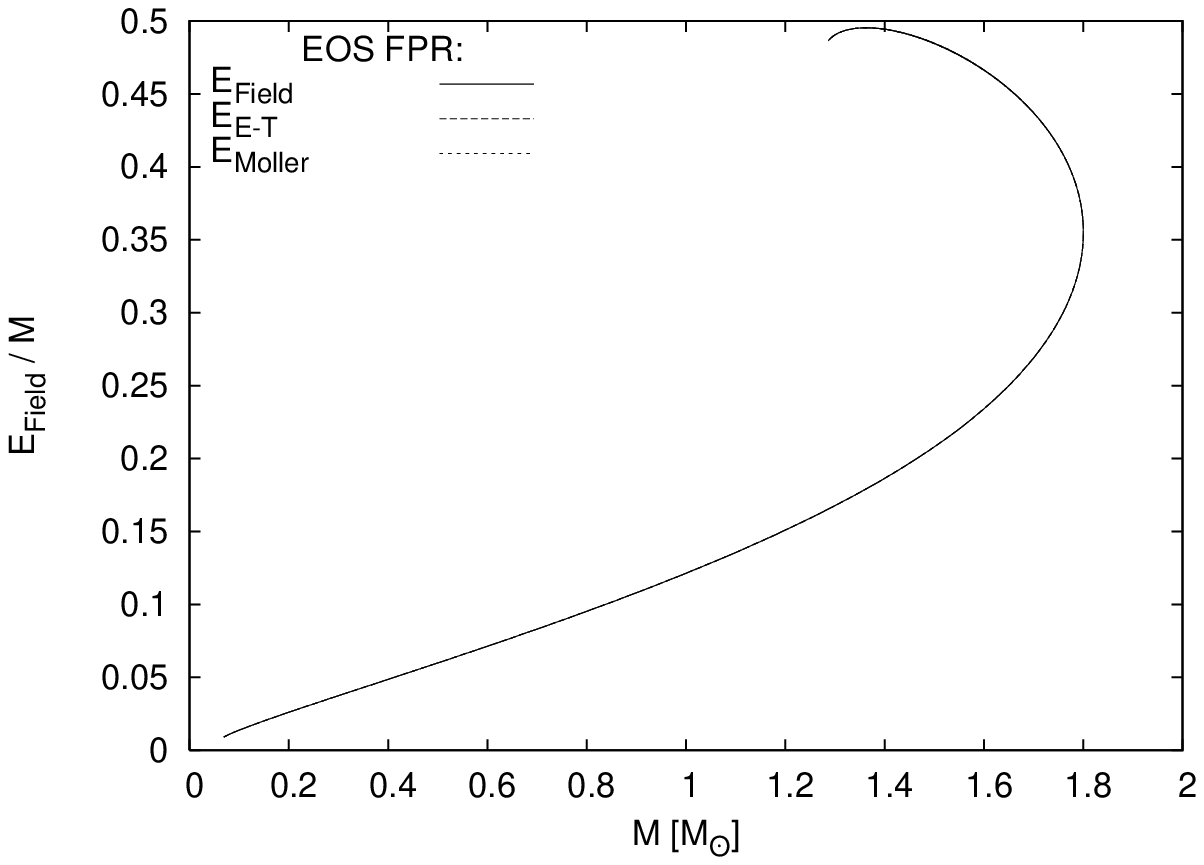}\includegraphics[height=4.3cm]{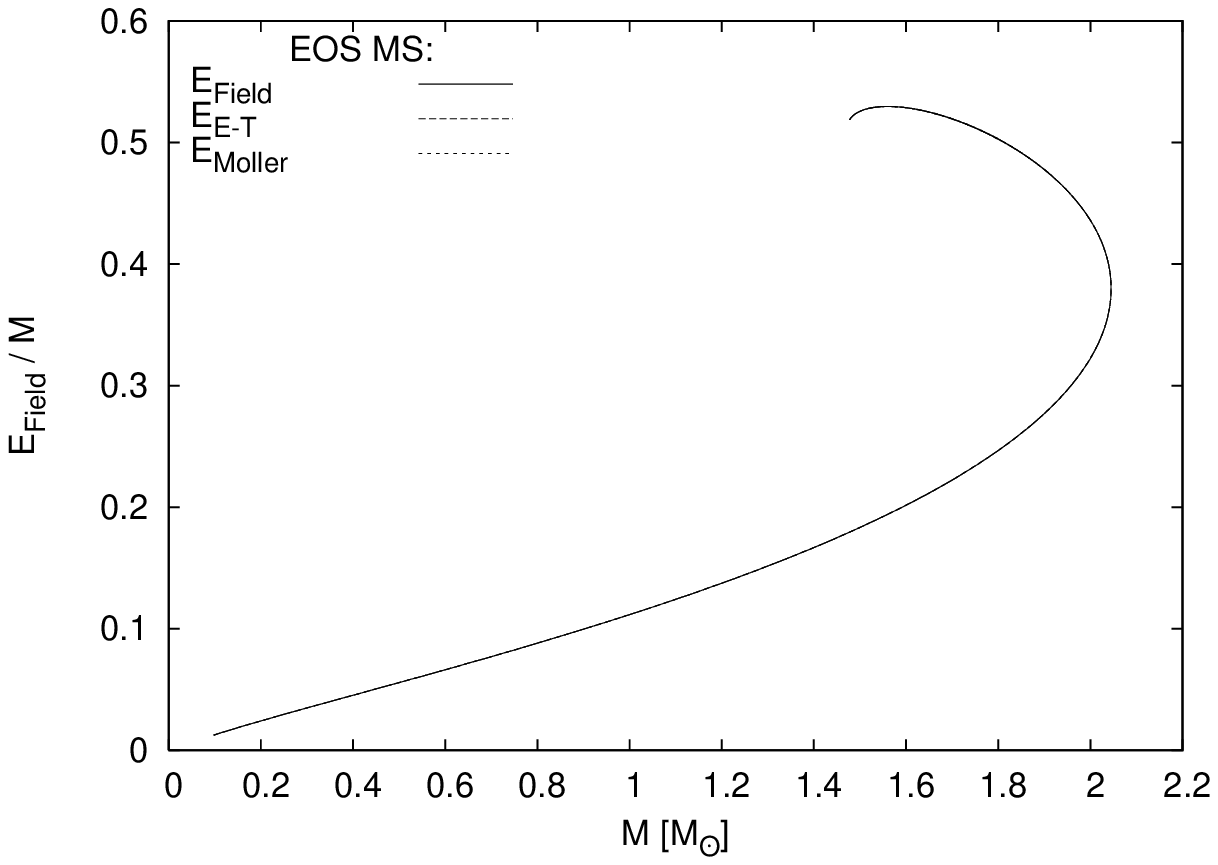}
	\end{center}
	\caption{Gravitational field energy of the neutron star. Curves, corresponding to all three definitions of the
	gravitational field energy, (9),(11) and (12) coincide.}
	\label{fig:ns_field_energy}
\end{figure}

In the following we calculate the graviational field energy contribution to the neutron star mass in terms of often used 
pseudotensors (which are discussed below in some details). From equation (\ref{e-t}) we have 
the field energy in the Einstein-Tolman prescription,
\begin{equation}
E^{E-T}_{Field}={4\pi G\over c^2} \int_0^R e^{{\nu/2}}\,M(r)\,{\left( {1 \over 1 -2\,G\,M(r)/c^2\,r} \right) }^{{3/2}}\,
    \left( P(r) + c^2\,\rho(r) \right) rdr.
\label{eg2}
\end{equation}
Using the  M\o ller pseudotensor(\ref{moller}) we find the gravitational field energy in the form
\begin{equation}
E^{Moller}_{Field}= 12\pi \int_0^R e^{{\nu/2}}\,P(r)\,{\sqrt{{1 \over 1-2\,G\,M(r)/c^2\,r}}}\, r^2dr.
\label{eg3}
\end{equation}
The definitions of the gravitational field energy of the neutron star Eqs.
(\ref{eg1}), (\ref{eg2}) and (\ref{eg3}) give the same value of the field 
 energy contribution. This holds for all equations of state we study as it is 
 shown in Fig. \ref{fig:ns_field_energy}. 

\begin{figure}[htb]
	\begin{center}
	
\includegraphics[height=4.3cm]{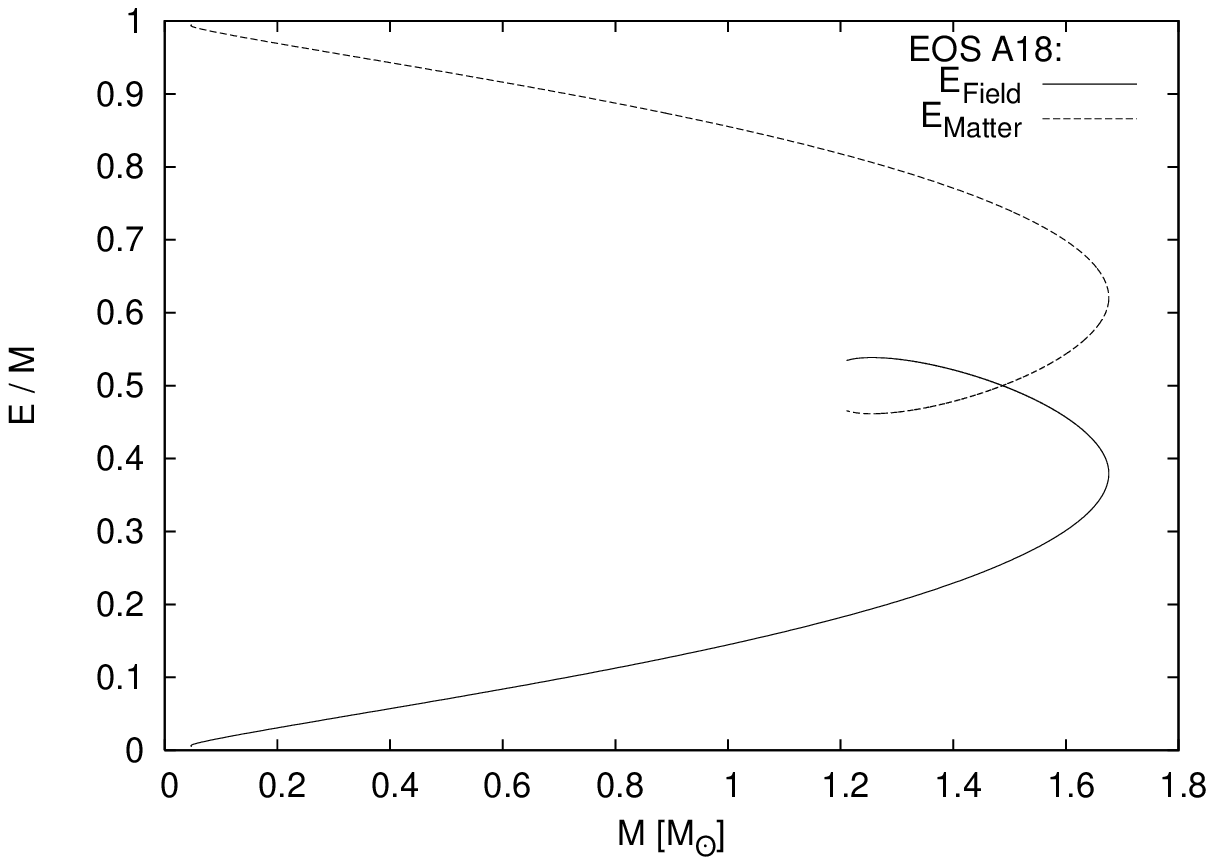}\includegraphics[height=4.3cm]{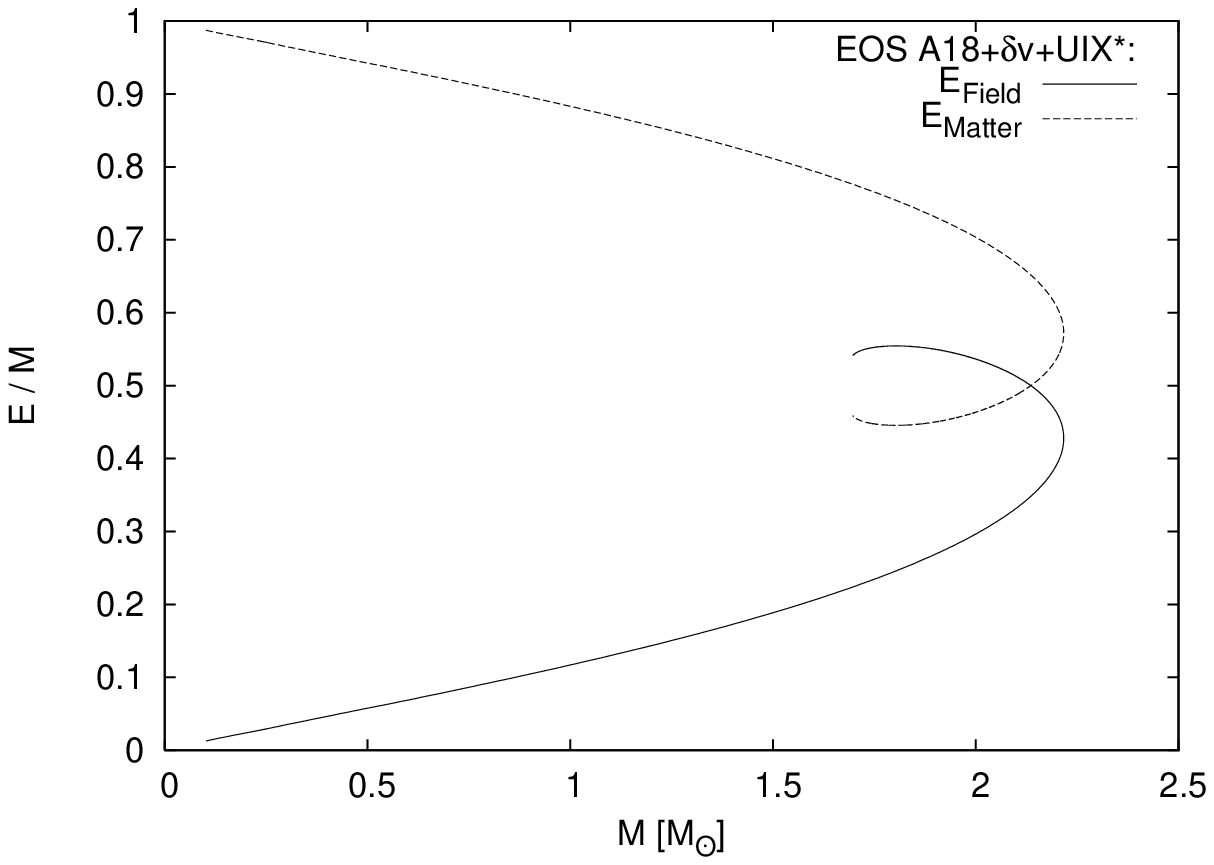}
	
\includegraphics[height=4.3cm]{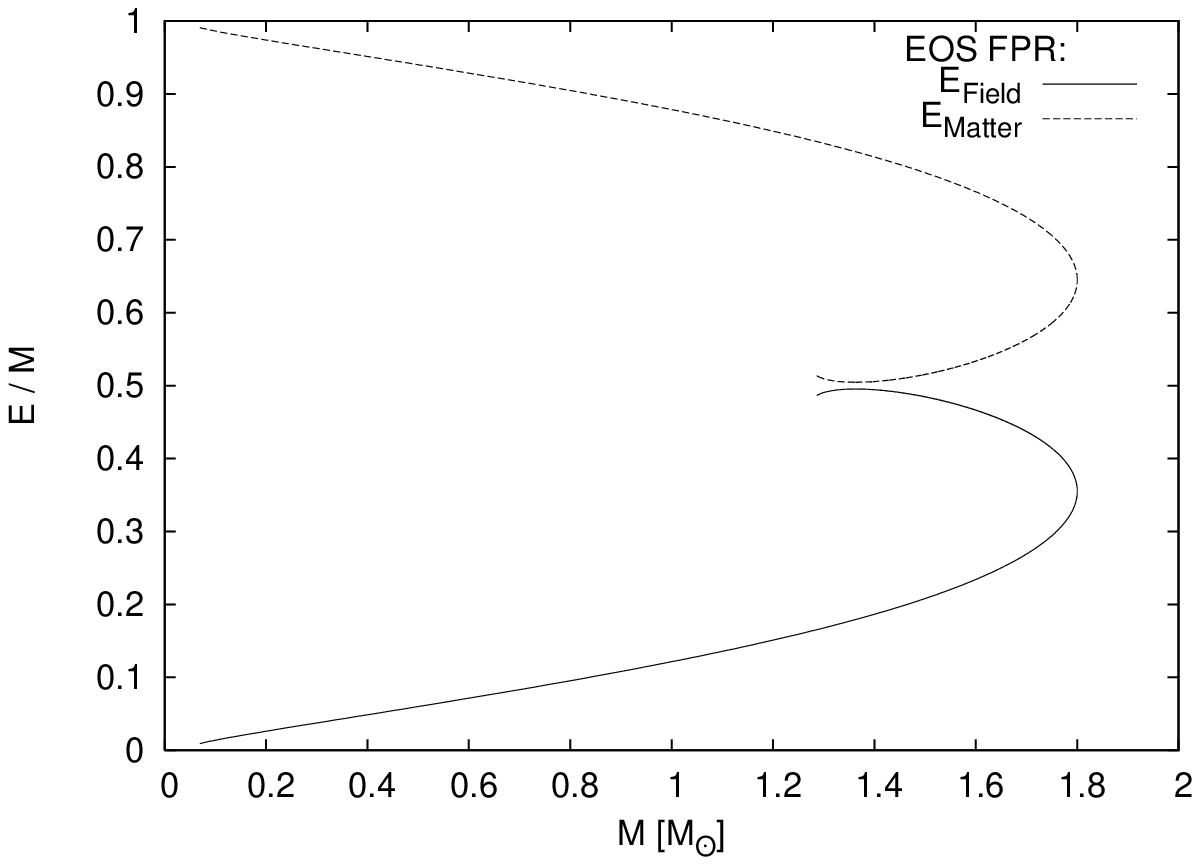}\includegraphics[height=4.3cm]{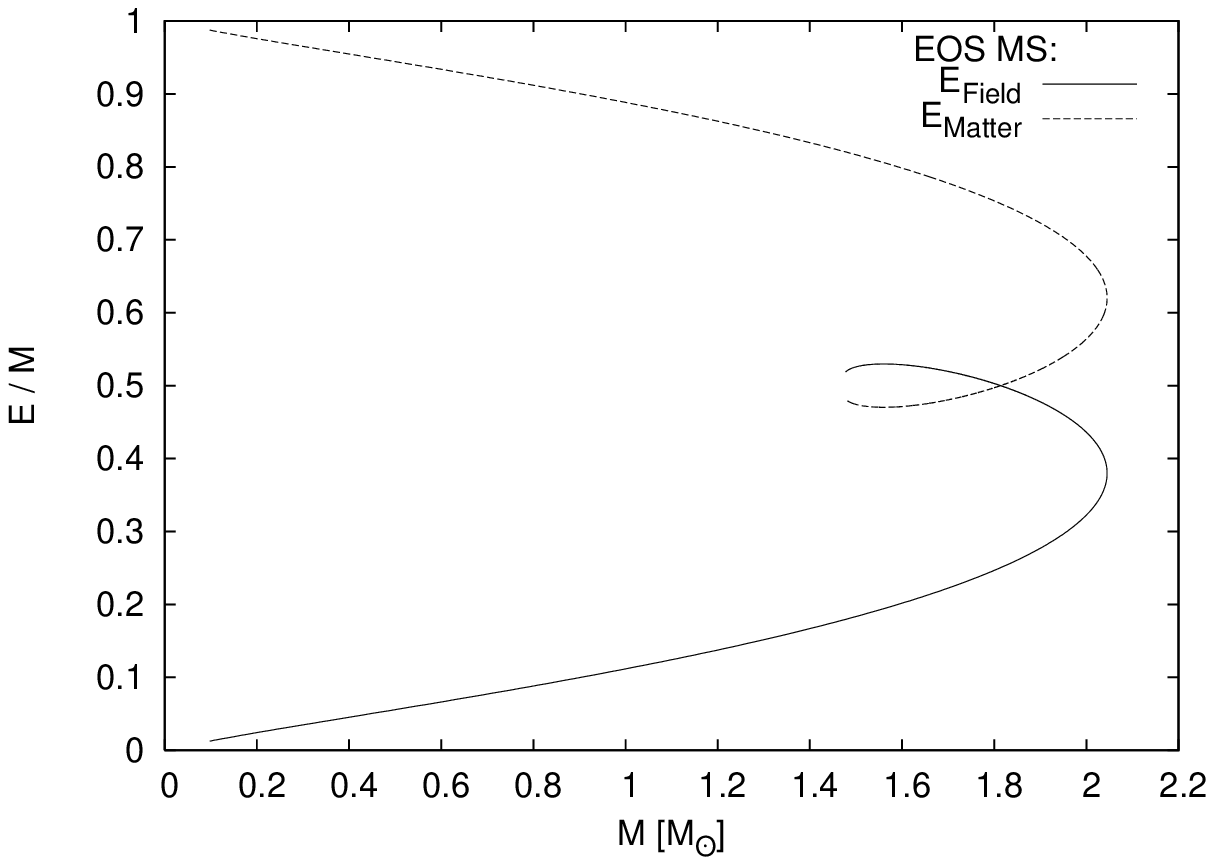}
	\end{center}
	\caption{Comparison of the gravitational field energy and matter energy of neutron star}
	\label{fig:ns_field_matter}
\end{figure}

Now we can compare  contributions of the matter energy and the gravitational field energy to neutron star 
masses. In Fig. \ref{fig:ns_field_matter} we show both energies for all equations of state we use. For low masses of 
neutron stars the matter energy dominates. For "canonical" neutron star mass $M=1.44M_{\odot}$ the gravitational field energy
contributes about 20\% to the total gravitational mass. This contribution increases significantly for maximum
mass neutron stars, exceeding 40\% for the A18+$\delta$v+UIX* EOS. Crossing point of the curves in Fig. \ref{fig:ns_field_matter} 
indicates the star for which half of the gravitational mass is comprised by  energy of its gravitational field. One can see 
that  this occurs for three equations of state, albeit for unstable stars.

For completeness we provide here some formulae for two pseudotensors we used in our calculations. We tried a
number of gravitational field pseudotensors, available in the literature. We tried Einstein-Tolman, M\o ller, Landau-Lifshitz
and Weinberg pseudotensors. However, not all of them were suitable for our calculations. 
To obtain the formulae given below we calculated all the components in Gallilean coordinates (see Appendix A) and employed 
the GRTensor program to perform the algebra.
Eventually, we performed numerical integration indicated in Eqs.(8, 9, 11, 12) for
Einstein-Tolman and M\o ller pseudotensors.

\subsection{Einstein-Tolman pseudotensor}
The energy-momentum pseudotensor of Einstein  and Tolman \cite{Tolman} is given by the formulae:
\begin{equation}
\theta^{\nu}_{\mu} = {c^4 \over 16\pi G} H^{\nu \gamma}_{\mu},_{\gamma},
\label{pseudo_ET}
\end{equation}
where $\theta^{\nu}_{\mu} = \sqrt{-g}(t^{\nu}_{\mu} + T^{\nu}_{\mu})$, and
\begin{equation}
H^{\nu \gamma}_{\mu} = {g_{\mu\rho} \over \sqrt{-g}} \left[-g\left(g^{\nu\rho}g^{\gamma \sigma}-g^{\gamma
\rho}g^{\nu\sigma}\right)\right],_{\sigma}.
\label{tensor_H}
\end{equation}
Here $g$ is the determinant of the metric tensor. 
This pseudotensor satisfies the local conservation laws:
\begin{equation}
{\partial \theta^{\nu}_{\mu} \over \partial x^{\nu}} =0.
\end{equation}
For the metric (\ref{metrics}) we find the energy density for the Einstein-Tolman pseudotensor to be
\begin{equation}
t^{0}_{0}={G\,M(r)\,\left( P(r) + c^2\,\rho(r)  \right) \over -2\,G\,M(r) + c^2\,r}.
\label{e-t}
\end{equation}
Some technical details of the calculations are shown in the Appendix B. The same result was obtained in Ref. \cite{DeHuaWen}.

\subsection{M\o ller pseudotensor}
The energy-momentum complex of M\o ller \cite{Moller} is given by the expression
\begin{equation}
M^{\nu}_{\mu} = {c^4\over 8\pi G}\,\chi^{\nu \gamma}_{\mu},_{\gamma}
\label{tensorM}
\end{equation}
where $M^{\nu}_{\mu} = \sqrt{-g}(t^{\nu}_{\mu} + T^{\nu}_{\mu})$ and the antisymmetric tensor $\chi^{\nu \gamma}_{\mu}$ is
\begin{equation}
\chi^{\nu \gamma}_{\mu} = \sqrt{-g}\left(g_{\mu \sigma},_{\rho}-g_{\mu \rho},_{\sigma}\right)g^{\nu \rho}g^{\gamma \sigma}.
\end{equation}
For metric (\ref{metrics}) we have :
\begin{equation}
t^{0}_{0}= 3\, P(r).
\label{moller}
\end{equation}
Some useful formulae are shown in the Appendix C.

\subsection{Landau pseudotensor}
The Landau pseudotensor reads \cite{LL}:
\begin{equation}
L^{\mu \nu} = {c^4 \over 16 \pi G} S^{\mu \nu \rho \sigma},_{\rho \sigma}
\end{equation}
where $S^{\mu \nu \rho \sigma} = -g(g^{\mu \nu}g^{\rho \sigma}-g^{\mu \rho}g^{\nu \sigma})$ and $L^{\mu \nu} = -g(T^{\mu \nu} + t^{\mu \nu})$.

The energy density is then given by
\begin{equation}
t^{0}_{0} = \frac{c^2\,G\,M(r)\,\left[ M(r) - 4\,\pi \,r^3\,\rho(r)  \right] }{2\,\pi \,r^3\,\left( 2\,G\,M(r) - c^2\,r \right) }
\end{equation}
One can notice that this energy density becomes negative outside the stellar radius. This is unphysical and we
discard the Landau pseudotensor in the following. The fact that the Landau pseudotensor gives the negative
energy density of the gravitational field of the Schwarzschild metric was first recognized by Virbhadra \cite{Virbhadra}.
One could probably attribute this behaviour to a peculiar dependence on $-g$, as observed by 
Cooperstock and Rosen \cite{Rosen}.  Some details of derivation of the $t_0^0$ component of the Landau-Lifshitz
pseudotensor, are given in Appendix D. 

As far as the Weinberg's pseudotensor is concerned, the relevant components vanish (Appendix E).

\section{The problem of localization of the gravitational field energy}
The fact that all three definitions of gravitational field energy of neutron stars give the same value is
reassuring. One can be quite confident that what we find is really the prediction of the general relativity
theory as to the gravitational energy field contribution to the neutron star mass.

\begin{figure}[ht]
	\begin{center}
	
\includegraphics[height=4.3cm]{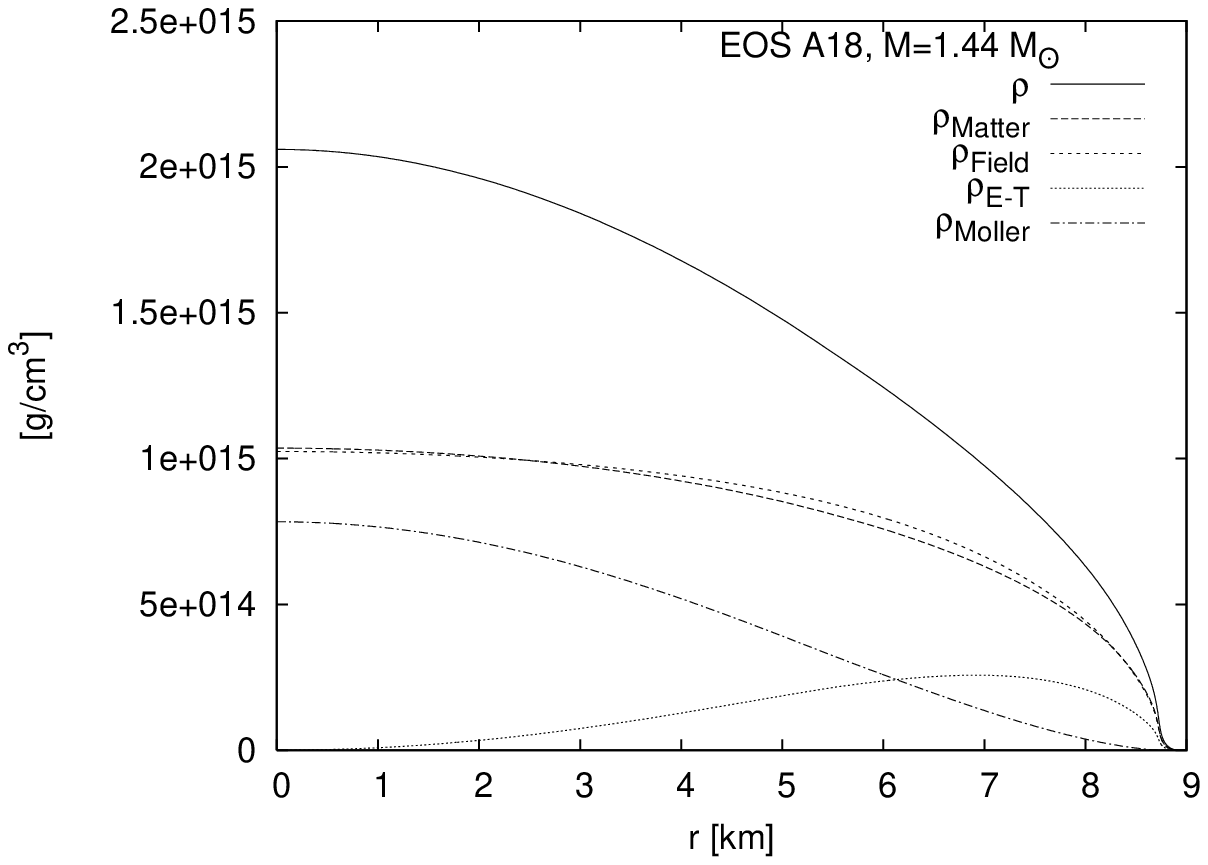}\includegraphics[height=4.3cm]{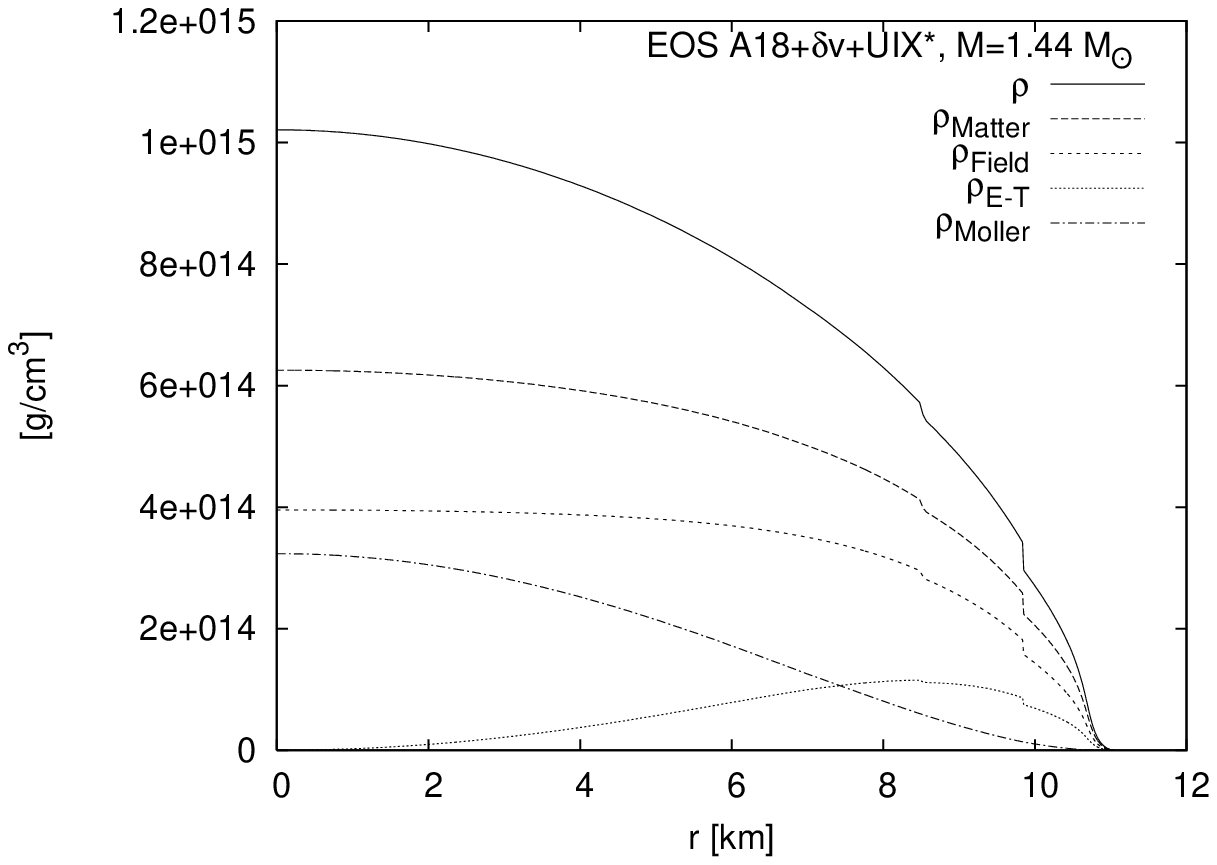}
	
\includegraphics[height=4.3cm]{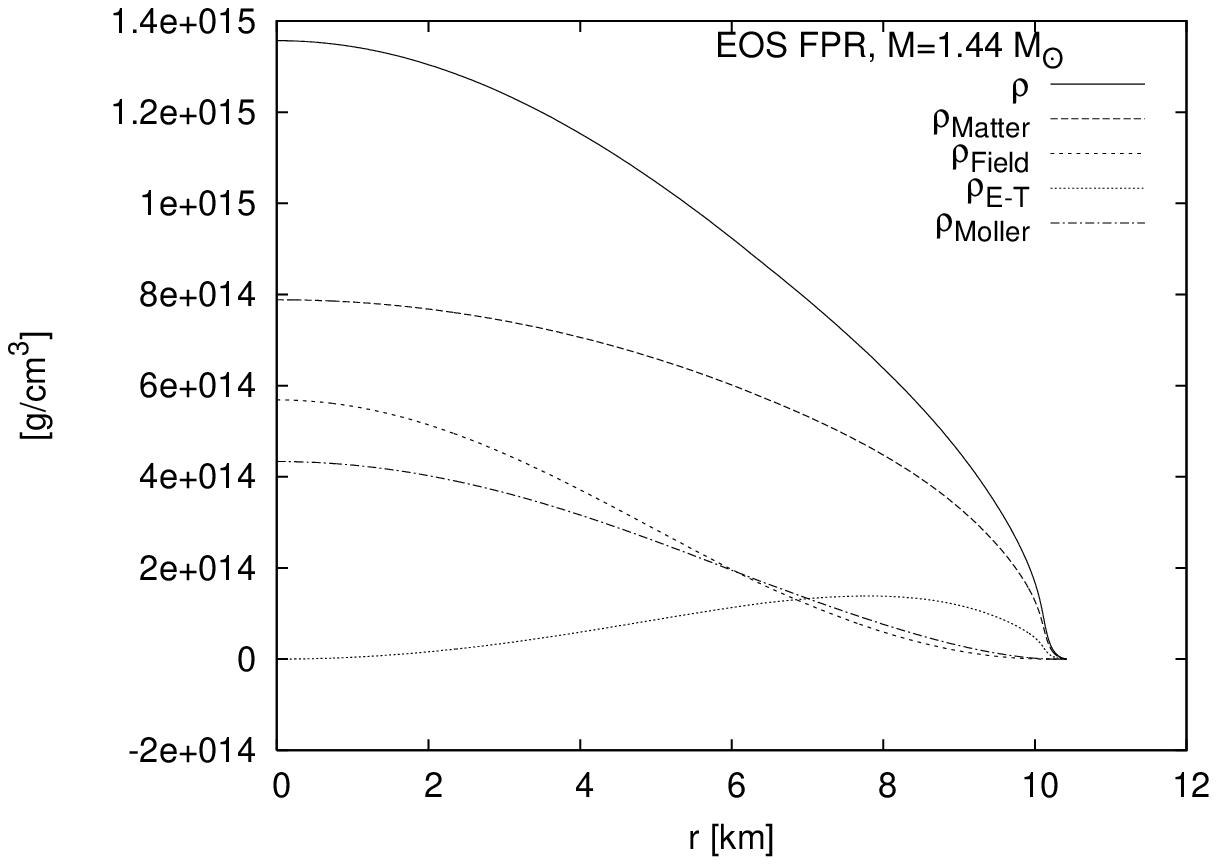}\includegraphics[height=4.3cm]{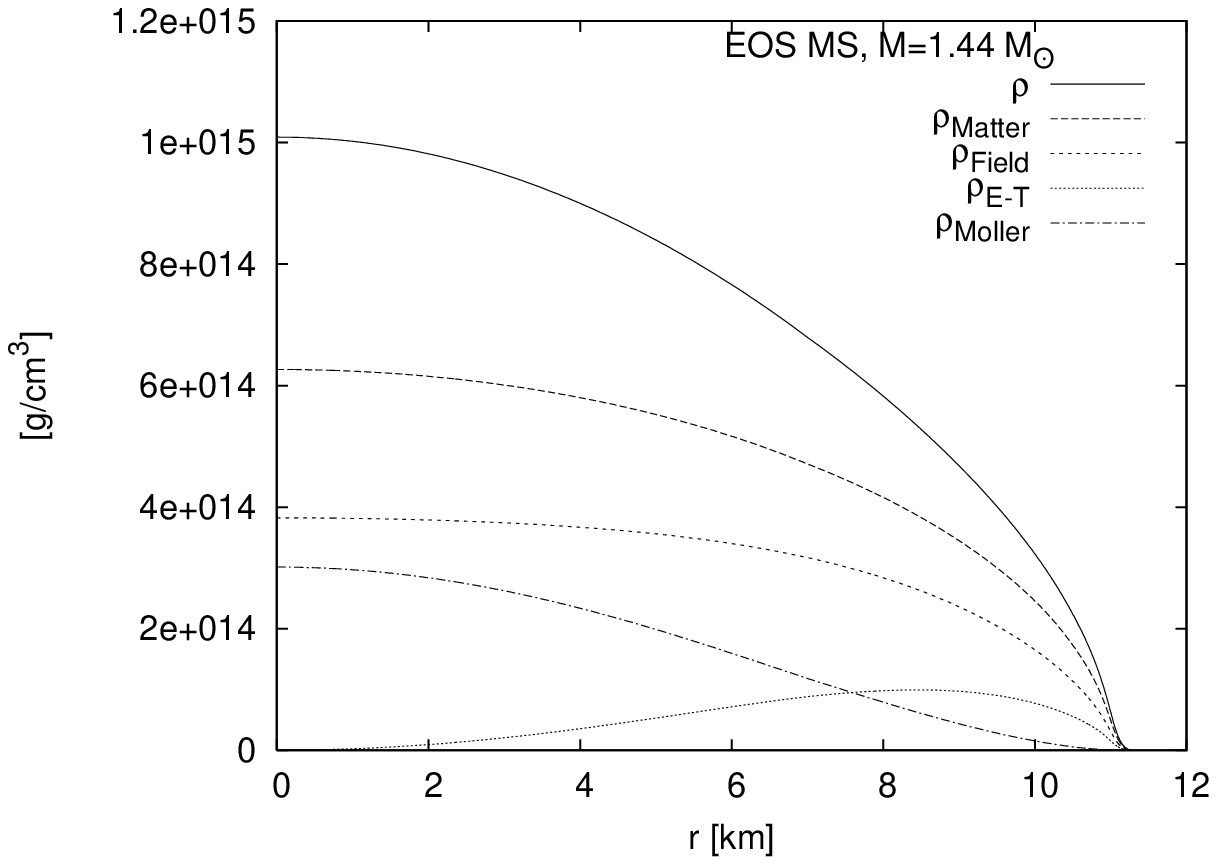}
	\end{center}
	\caption{The energy density of the neutron star gravitational field according to different definitions  for  
	the ''canonical'' mass neutron star $1.44\,M_{\odot}$.}
	\label{fig:density_canonical}
\end{figure}

\begin{figure}[hbt]
	\begin{center}
	
\includegraphics[height=4.3cm]{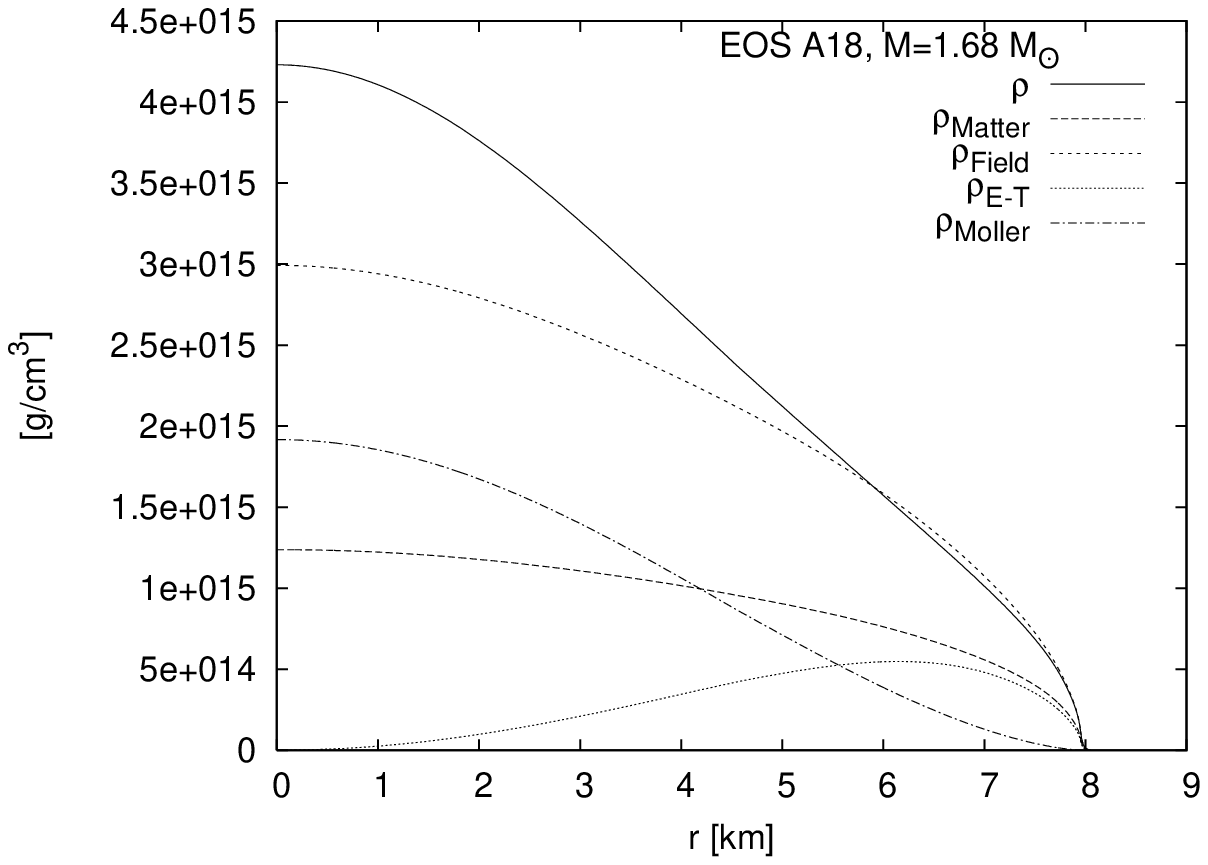}\includegraphics[height=4.3cm]{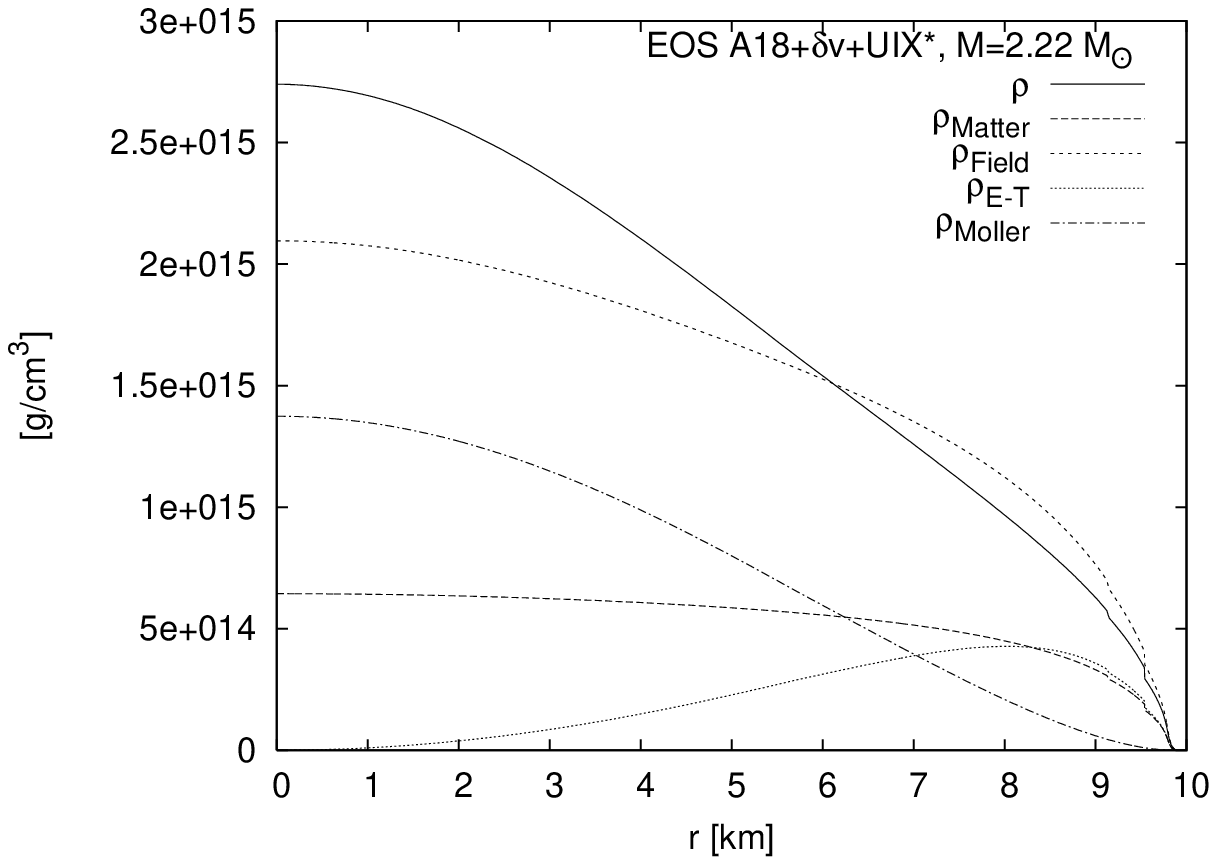}
	
\includegraphics[height=4.3cm]{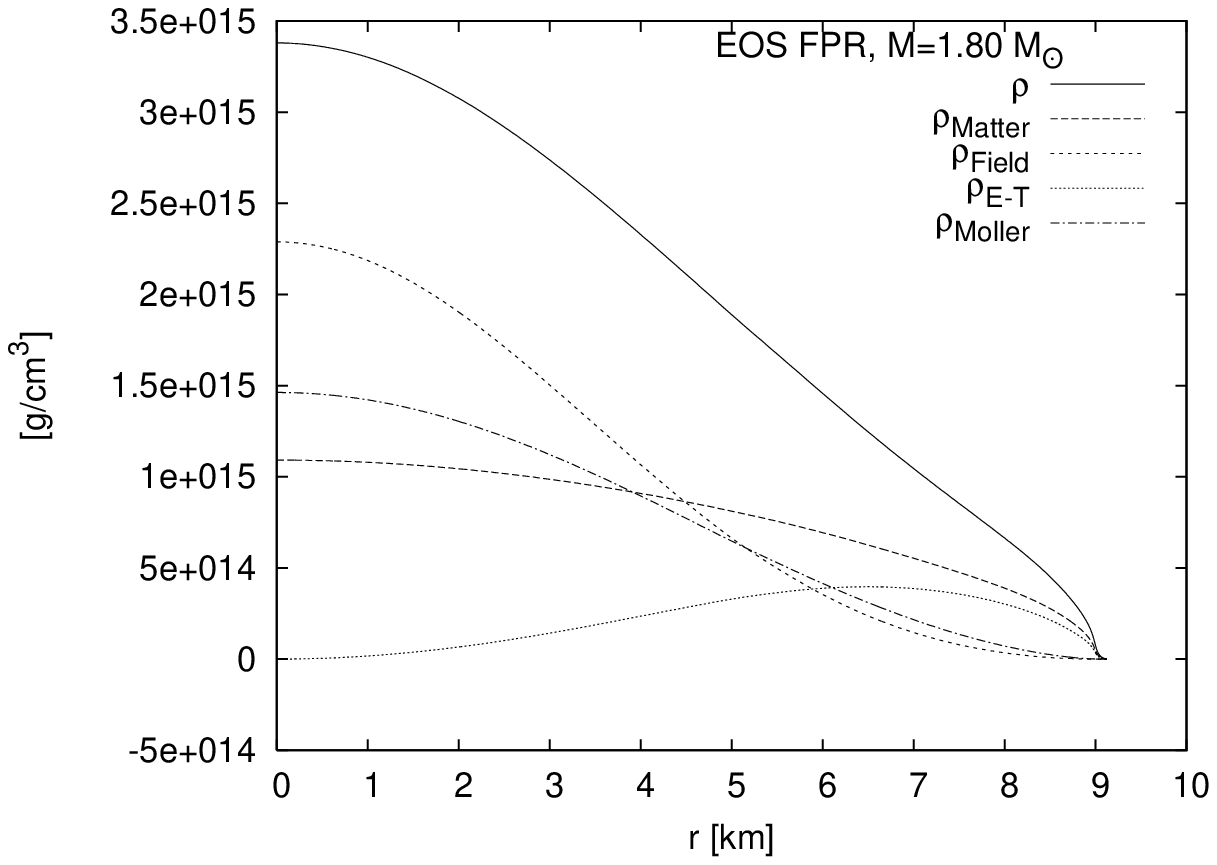}\includegraphics[height=4.3cm]{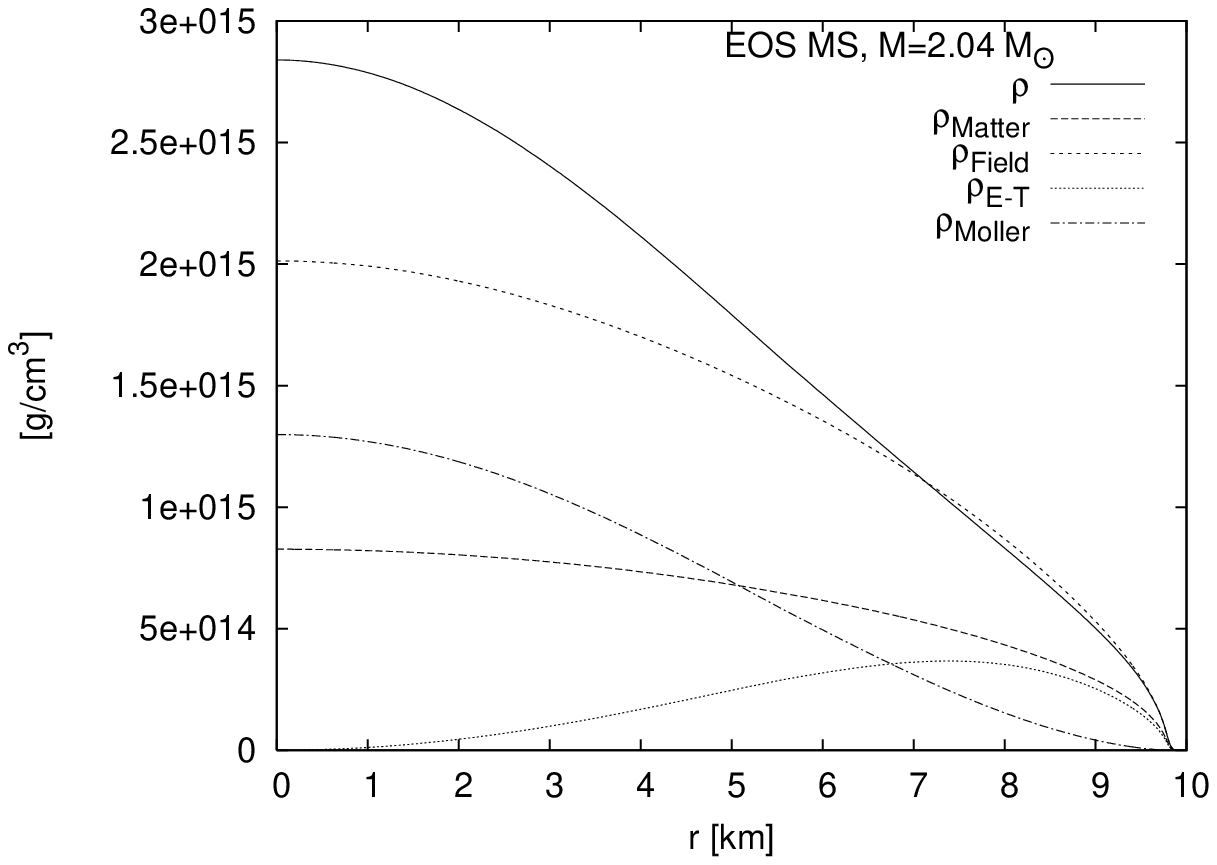}
	\end{center}
	\caption{The same as in Fig.5 for maximum mass neutron stars}
	\label{fig:density_max}
\end{figure}

There is another important conclusion to be drawn: In all cases we consider the gravitational field energy is
rather well localized. To show this we plot in Figs. \ref{fig:density_canonical} and \ref{fig:density_max} "gravitational field energy density" for all three definitions we use. Although these are not to be understood as physically meaningful, plots in Figs. \ref{fig:density_canonical} and \ref{fig:density_max} show they share the same feature: namely all  densities are localized within the star. 

The first density corresponds to our definition of the field energy (\ref{eg1}) and is given by Eq.(10).
The other two  gravitational field energy densities correspond to Einstein-Tolman pseudotensor,
\begin{equation}
\rho^{E-T}_{Field} = {e^{{\nu/2}}G\,M(r)\,\left( P(r) + c^2\,\rho(r)  \right) \over -2\,G\,M(r) + c^2\,r}
\end{equation}
and to the M\o ller pseudotensor,
\begin{equation}
\rho^{Moller}_{Field} = 3\,e^{{\nu/2}}P(r)
\end{equation}

In Figs. \ref{fig:density_canonical}, \ref{fig:density_max} we show results for "canonical" neutron star and for maximum 
mass neutron stars respectively.

The fact that the whole gravitational field energy is accounted for within the star radius is fully consistent
with the findings of Nahmad-Achar and Schutz \cite{NAS} who found that the energy associated with the Schwarzschild
space-time was independent of radius for radii exceeding the radius of spherically symmetric mass. 
We thus arrive at the conclusion that for spherically symmetric compact objects with Einstein-Tolman and M\o ller 
gravitational field pseudotensors, the whole gravitational mass is localized within the star, including the
gravitational field contribution which for neutron stars is substantial - reaching 40\% of the total mass for
maximum mass neutron stars. 

In this case the total gravitational mass, $M$, which is the mass of the exterior Schwarzschild metric has all the contributions
inside the stellar radius, and thus can be used for calculations of such properities as the last stable orbit around the neutron
star, for sufficiently massive stars. However, there is some unusual property of the gravitational field in this approach: the 
field just below the surface of the star has the energy which contributes to the stellar mass, whereas above the
stellar surface the field  does not contribute to the neutron star mass: no gravitational field energy is localized 
outside the stellar surface.  

\section{Equations of state and nuclear models}\label{EOSNM}
In this section equations of state of dense matter, used for calculating curves in the above plots, are briefly discussed. 

The density dependence of the pressure $P(\rho)$ plays the crucial role in calculations
of  neutron star models.
The mass density of matter, $\rho$, includes kinetic and interaction energy density of the nucleons
and rest masses of neutrons and protons
$\rho(n_B) =\varepsilon\left( n_N,n_P\right) +n_Nm_N+n_Pm_P$,
while the pressure of the nuclear fluid is 
\begin{equation}
P\left( n_B\right) =n_B^2\frac{d\left( \varepsilon/n_B\right) }{dn_B}.
\end{equation}
The baryon number density $n_B=n_N+n_P$ is the sum of the neutron and proton number densities.

We chose four realistic nuclear interaction models which give us the energy density as a function of the baryon number
density, $\epsilon(n_B)$.
These are interactions derived by Myers and Swiatecki \cite{MySwZSP} (MS),
the Friedman and Pandharipande interaction model \cite{FrPaNP} (as parametrized by Ravenhall in Ref.\cite{LaARNPS}) (FPR)
and two modern potential models, A18 and A18+$\delta$v+UIX* \cite{Akm/Pan/Rav PR C58},
which provide a good fit to the nucleon-nucleon scattering data in the Nijmegen data base.

Myers and Swiatecki proposed a nuclear model based on the Thomas-Fermi approximation \cite{My/Sw AP204,My/Sw AP211}.
They extended the model of Seyler and Blanchard \cite{Se/Bl PR124} by improving
four-parameter interaction force between nucleons.
By fitting two additional parameters they obtained an excellent agreement with the ground state energies
of about 1600 atomic nuclei.
The energy density reads

\begin{eqnarray}
\lefteqn{\varepsilon_{MS}\left( n_N,n_P\right)  =  A\left( n_N^{\frac53}+n_P^{\frac53}\right)-B_1\left( n_N^2+n_P^2\right)}
\nonumber \\ & &
-2B_2n_Nn_P + C_1\left( n_N^{\frac83}+n_P^{\frac83}\right)\nonumber \\ & &
+C_2\left( n_N^{\frac53}n_P+n_Nn_P^{\frac53}\right) -D\left( 5n_N^{\frac23}n_P-n_P^{\frac53}\right),
\end{eqnarray}
where the parameter values are:
\begin{center}
\begin{tabular}{l l}
$A=88.247\,MeV\,fm^2$ & $B_1=169.217\,MeV\,fm^3$  \\
$B_2=286.715\,MeV\,fm^3$ & $C_1=358.664\,MeV\,fm^5$ \\
$C_2=1167.630\,MeV\,fm^5$ & $D=50.148\,MeV\,fm^2$.\\
\end{tabular}
\end{center}

D. G. Ravenhall has found a simple formula to reproduce nuclear matter calculations of Friedman and Pandharipande \cite{FrPaNP}.
The interaction model fits the s-, p-, d-, and f-wave phase shifts in the low density limit,
and at high density the hypernetted chain calculations which makes the effective interactions
similar to a contact two-body force.
Adjustable tensor force contribution is also included to obtain good saturation properties.
These model forces have been fitted to the nuclear matter calculations at low and high densities. The energy density is

{\setlength\arraycolsep{2pt}
\begin{eqnarray}
\lefteqn{\varepsilon_{FPR}\left( n_N,n_P\right)  =  \left( \frac{1}{2m_N}+B_N\right) \tau_N
+\left( \frac{1}{2m_P}+B_P\right) \tau_P \nonumber} \\ & &
+n_B^2\left[ a_1+a_2e^{-b_1n_B}+\left( \frac12-x\right) ^2
\left( a_3+a_4e^{-b_1n_B}\right) \right] \nonumber \\ & &
+n_Be^{-b_2n_B^2}\left[ a_5+a_6n_B+\left( \frac12-x\right) ^2\left( a_7+a_8n_B\right) \right] \textrm{,}
\end{eqnarray}}
where:
\begin{center}
\begin{tabular}{l l }
$n_B=n_N+n_P$, & $x=\frac{n_P}{n_B}$, \\
$B_i=\left( a_9n_B+a_{10}n_i\right) e^{-b_3n_B}$, & $\tau_i=\frac35\left( 3\pi^2\right) ^{\frac23}n_i^{\frac53}$ \\
\end{tabular}
\end{center}
for $i=N\textrm{,} \, P$.
The parameter values are:
\begin{center}
\begin{tabular}{l l l}
$a_1=1054.0\,MeV\,fm^3$ & $a_2=-1393.0\,MeV\,fm^3$ & $a_3=-2316.0\,MeV\,fm^3$  \\
$a_4=2859.0\,MeV\,fm^3$ & $a_5=-1.78\,MeV$ & $a_6=-52.0\,MeV\,fm^3$ \\
$a_7=5.5\,MeV$ & $a_8=197.0\,MeV\,fm^3$ & $a_9=89.8\,MeV\,fm^5$ \\
$a_{10}=-59.0\,MeV\,fm^5$ & $b_1=0.284\,fm^3$ & $b_2=42.25\,fm^6$ \\
$b_3=0.457\,fm^3$ & & .\\
\end{tabular}
\end{center}

Akmal, Pandharipande and Ravenhall used variational chain summation methods and created
new Argonne $v_{18}$ two-nucleon interaction which fits all of the nucleon-nucleon scattering data
in the Nijmegen data base.

The A18+$\delta v$+UIX* model comprises the boost correction $\delta v$ to the two-nucleon interaction,
which gives the leading relativistic effect of order $\left( v/c\right) ^2$
and the three-nucleon interaction Urbana model IX* which predicts a transition in nucleon matter
to a phase with neutral pion condensate at the density $0.195\,fm^{-3}$ for pure neutron matter
and $0.32\,fm^{-3}$ for symmetric nuclear matter.

The energy density for both A18 and A18+$\delta v$+UIX* models can be given by a common formula,

{\setlength\arraycolsep{2pt}
\begin{eqnarray*}
\lefteqn{\varepsilon_{APR}\left( n_N,n_P\right)    = 
\left( \frac1{2m_N}+B_N\right) \tau_N+\left( \frac1{2m_P}+B_P\right) \tau_P} \nonumber \\ & &
   -4n_Nn_P\left( p_1+p_2n_B+p_6n_B^2
+\left( p_{10}+p_{11}n_B\right) e^{-p_9^2n_B^2}\right) \nonumber \\ & &
   -\left(n_N-n_P\right) ^2\left( p_{12}/n_B+p_7+p_8n_B+p_{13}e^{-p_9^2n_B^2}\right)
\end{eqnarray*}
for A18+$\delta v$+UIX* with $n_B>p_{19}$ or $n_B>p_{20}$ additionaly:
\begin{eqnarray}
\lefteqn{ -4n_Nn_P\left(n_B-p_{19}\right) \left( p_{17}+p_{21}\left(n_B-p_{19}\right) \right)
e^{p_{18}\left(n_B-p_{19}\right) } \nonumber} \\ & &
  -\left(n_N-n_P\right) ^2\left(n_B-p_{20}\right) \left( p_{15}+p_{14}\left(n_B-p_{20}\right) \right)
e^{p_{16}\left(n_B-p_{20}\right) } 
\end{eqnarray}}
where: $B_i=\left( p_3n_B+p_5n_i\right) e^{-p_4n_B}$,
$\tau_i=\frac35\left( 3\pi^2\right) ^{\frac23}n_i^{\frac53}$ for $i=N\textrm{,} \, P$.
The parameters $p_3=89.8\,MeV\,fm^5$, $p_4=0.457\,fm^3$ and $p_5=-59.0\,MeV\,fm^5$ are common
for these two models.

For A18 model the parameters are:
$p_1=297.6\,MeV\,fm^3$, $p_2=-134.6$ $MeV\,fm^6$, $p_6=-15.9\,MeV\,fm^9$, $p_7=215.0\,MeV\,fm^3$,
$p_8=-116.5$ $MeV\,fm^6$, $p_9=6.42\,fm^3$, $p_{10}=51\,MeV\,fm^3$, $p_{11}=-35\,MeV\,fm^6$,
$p_{12}=-0.2\,MeV$ and $p_{13}=0\,MeV\,fm^3$.

For A18+$\delta v$+UIX* model the parameters are:
$p_1=337.2\,MeV\,fm^3$, $p_2=-382\,MeV\,fm^6$, $p_6=-19.1\,MeV\,fm^9$, $p_7=214.6\,MeV\,fm^3$,
$p_8=-384\,MeV\,fm^6$, $p_9=6.4\,fm^3$, $p_{10}=69\,MeV\,fm^3$, $p_{11}=-33\,MeV\,fm^6$,
$p_{12}=0.35\,MeV$, $p_{13}=0\,MeV\,fm^3$, $p_{14}=0\,MeV\,fm^9$, $p_{15}=287\,MeV\,fm^6$,
$p_{16}=-1.54\,fm^3$, $p_{17}=175.0\,MeV\,fm^6$, $p_{18}=-1.45\,fm^3$, $p_{19}=0.32\,fm^{-3}$,
$p_{20}=0.195\,fm^{-3}$ and $p_{21}=0\,MeV\,fm^9$.

\appendix
\section{Metric in Cartesian coordinates}

First we change the spherical coordinates to the Cartesian ones. Using transformation 
\[\begin{array}{cll}
r  & = &   \sqrt{x^2 + y^2 + z^2} \\
\phi & = &  \arctan (y/x) \nonumber \\
\theta & = &  \arccos (z/\sqrt{x^2 + y^2 + z^2}) 
\end{array}\]
we obtain the metric in Cartesian coordinates,

\begin{eqnarray}
\nonumber
\lefteqn{
ds^2 = e^{\nu}c^2dt^2 - \frac{e^{\lambda}x^2+y^2+z^2}{x^+y^2+z^2}dx^2
-\frac{x^2+e^{\lambda}y^2+z^2}{x^+y^2+z^2}dy^2}
\\
\nonumber
&-&\frac{x^2+y^2+e^{\lambda}z^2}{x^+y^2+z^2}dz^2 - \frac{(e^{\lambda}-1)2yz}{x^+y^2+z^2}dy dz+
\\
&+& \left(-\frac{(e^{\lambda}-1)2xy}{x^+y^2+z^2}dy - \frac{(e^{\lambda}-1)2xz}{x^+y^2+z^2}dz\right)dx
\label{metryka_kart}
\end{eqnarray}

The components of the covariant metric tensor are
\begin{equation}
g_{\mu \nu} =	\left(\begin{array}{cccc}
e^{\nu} & 0 & 0 & 0 \\
0& - \frac{e^{\lambda }\,x^2 + y^2 + z^2}{x^2 + y^2 + z^2} & -\frac{\left( -1 + e^{\lambda } \right) \,x\,y}{x^2 + y^2 + z^2} & -\frac{\left( -1 + e^{\lambda } \right) \,x\,z}{x^2 + y^2 + z^2} \\
0 & -\frac{\left( -1 + e^{\lambda } \right) \,x\,y}{x^2 + y^2 + z^2} & -\frac{x^2 + e^{\lambda }\,y^2 + z^2}{x^2 + y^2 + z^2} & - \frac{-\left( y\,z \right)  + e^{\lambda }\,y\,z}{x^2 + y^2 + z^2}\\
0 & -\frac{\left( -1 + e^{\lambda } \right) \,x\,z}{x^2 + y^2 + z^2} & - \frac{-\left( y\,z \right)  + e^{\lambda }\,y\,z}{x^2 + y^2 + z^2} & \frac{x^2 + y^2 + e^{\lambda }\,z^2}{x^2 + y^2 + z^2}
\end{array}\right)
\end{equation}
The determinant of this metric tensor is $g = -e^{\lambda  + \nu }$. Then we find the components of the contravariant metric 
tensor in Cartesian coordinates to be
\begin{equation}
g^{\mu \nu} =	\left(\begin{array}{cccc}
e^{-\nu} & 0 & 0 & 0 \\
0 & - \frac{x^2 + e^{\lambda }\,\left( y^2 + z^2 \right) }{e^{\lambda }\,\left( x^2 + y^2 + z^2 \right) } & \frac{\left( -1 + e^{\lambda } \right) \,x\,y}{e^{\lambda }\,\left( x^2 + y^2 + z^2 \right) } & \frac{\left( -1 + e^{\lambda } \right) \,x\,z}{e^{\lambda }\,\left( x^2 + y^2 + z^2 \right) } \\
0 & \frac{\left( -1 + e^{\lambda } \right) \,x\,y}{e^{\lambda }\,\left( x^2 + y^2 + z^2 \right) } & - \frac{y^2 + e^{\lambda }\,\left( x^2 + z^2 \right) }{e^{\lambda }\,\left( x^2 + y^2 + z^2 \right) } & \frac{\left( -1 + e^{\lambda } \right) \,y\,z}{e^{\lambda }\,\left( x^2 + y^2 + z^2 \right) } \\
0 & \frac{\left( -1 + e^{\lambda } \right) \,x\,z}{e^{\lambda }\,\left( x^2 + y^2 + z^2 \right) } & \frac{\left( -1 + e^{\lambda } \right) \,y\,z}{e^{\lambda }\,\left( x^2 + y^2 + z^2 \right) }& -\frac{e^{\lambda }\,\left( x^2 + y^2 \right)  + z^2}{e^{\lambda }\,\left( x^2 + y^2 + z^2 \right) }
\end{array}\right)
\end{equation}
Note, that now $\lambda$ and $\nu$ are functions of Cartesian coordinates $x$, $y$ and $z$.

\section{Calculation of the Einstein-Tolman pseudotensor}

We are interested in the energy component of the Einstein-Tolman pseudotensor \cite{Tolman}. Using 
the equations (\ref{pseudo_ET}) and (\ref{tensor_H}) one can calculate the nonvanishing components of $H_0^{0 \alpha}$

{\setlength\arraycolsep{2pt}
\begin{eqnarray}
\nonumber
H_{0}^{01} = e^{(\nu-\lambda)/2}\left(\partial_x(e^\lambda g^{11}) + \partial_y(e^\lambda g^{12}) + \partial_z(e^\lambda g^{13})  \right) \\
\nonumber 
  H_{0}^{02} = e^{(\nu-\lambda)/2}\left(\partial_x(e^\lambda g^{21}) + \partial_y(e^\lambda g^{22}) + \partial_z(e^\lambda g^{23})  \right)\\
  H_{0}^{03} = e^{(\nu-\lambda)/2}\left(\partial_x(e^\lambda g^{31}) + \partial_y(e^\lambda g^{32}) + \partial_z(e^\lambda g^{33})  \right)
\end{eqnarray}}
Inserting the componets of the contravariant metric tensor we have
{\setlength\arraycolsep{2pt}
\begin{eqnarray}
\nonumber
H_{0}^{01} = \frac{2\,\left(e^{\lambda } -1 \right) \,x}{x^2 + y^2 + z^2} \\
\nonumber
H_{0}^{02} = \frac{2\,\left(e^{\lambda } -1 \right) \,y}{x^2 + y^2 + z^2}\\
H_{0}^{03} = \frac{2\,\left(e^{\lambda } -1 \right) \,z}{x^2 + y^2 + z^2}
\end{eqnarray}
The 00 componet of $\theta$ pseudotensor is given by
\begin{equation}
\theta_0^0 =\frac{c^4}{16\pi G}\left(\partial_x H_{0}^{01}+\partial_y H_{0}^{02}+\partial_z H_{0}^{03}\right)
\end{equation}}
Taking the partial derivatives of $H_0^{0 i}$ we obtain

\begin{equation}
\theta_0^0 = \frac{e^{(\lambda  + \nu )/2}\,\left[ \left( -1 + e^{\lambda } \right) \,P(r) + 
      c^2\,\left( 1 + e^{\lambda } \right) \,\rho(r)  \right] }{2}
\end{equation}
Notice that $e^{-\lambda}=1-2GM(r)/c^2r$, so the 00 component of the Eintein-Tolman pseudotensor  is
\begin{equation}
\theta_0^0 =  \frac{e^{\frac{\nu }{2}}\,{\left( \frac{c^2\,r}{-2\,G\,M(r) + c^2\,r} \right) }^{\frac{3}{2}}\,
    \left( c^4\,r\,\rho(r)  + G\,M(r)\,\left( P(r) - c^2\,\rho(r)  \right)  \right) }{c^2\,r}
\end{equation}
Using definition of $\theta$ (\ref{pseudo_ET}) we obtain the equation (\ref{e-t}) which is the gravitational field 
energy density in the Einstein-Tolman prescription. The same result was obtained by \cite{DeHuaWen}.
\begin{equation}
t^{0}_{0}={G\,M(r)\,\left( P(r) + c^2\,\rho(r)  \right) \over -2\,G\,M(r) + c^2\,r}.
\end{equation}

\section{M\o ller prescription}
For metric in Cartesian coordinates the $\chi$ potential is
\begin{equation}
\chi_{0}^{0\gamma} = -\sqrt{-g}\left(\frac{\partial g_{00}} {\partial x^{\rho}}  g^{00} g^{\gamma\rho}\right)
\end{equation}
The nonzero componets of $\chi$ are
{\setlength\arraycolsep{2pt}
\begin{eqnarray}
\nonumber
\chi_{0}^{01} = - e^{(\lambda + \nu)/2}\frac{\nu'}{\sqrt{x^2+y^2+z^2}}\left(x\,g^{11} + y\,g^{12} + z\,g^{13}\right) \\
\nonumber
\chi_{0}^{02} = - e^{(\lambda + \nu)/2}\frac{\nu'}{\sqrt{x^2+y^2+z^2}}\left(x\,g^{21} + y\,g^{22} + z\,g^{23}\right) \\
\chi_{0}^{03} = - e^{(\lambda + \nu)/2}\frac{\nu'}{\sqrt{x^2+y^2+z^2}}\left(x\,g^{31} + y\,g^{32} + z\,g^{33}\right)
\end{eqnarray}}
where prime denotes derivative of the function $\nu$ with respect to $r=\sqrt{x^2+y^2+z^2}$. Inserting the components of 
the contravariant metric tensor and using the equation (\ref{me2}) we have

{\setlength\arraycolsep{2pt}
\begin{eqnarray}
\nonumber
\chi_{0}^{01} = \frac{e^{\frac{-\lambda  + \nu }{2}}\,x\,\left( c^4\,\left( -1 + e^{\lambda } \right)  + 
      8\,\pi\,G\,e^{\lambda }\,P\,\left( x^2 + y^2 + z^2 \right)  \right) }{c^3\,\left( x^2 + y^2 + z^2 \right) } \\
\nonumber
\chi_{0}^{02} = \frac{e^{\frac{-\lambda  + \nu }{2}}\,y\,\left( c^4\,\left( -1 + e^{\lambda } \right)  + 
      8\pi\,G\,e^{\lambda }\,P\,\left( x^2 + y^2 + z^2 \right)  \right) }{c^3\,\left( x^2 + y^2 + z^2 \right) } \\
\chi_{0}^{03} = \frac{e^{\frac{-\lambda  + \nu }{2}}\,z\,\left( c^4\,\left( -1 + e^{\lambda } \right)  + 
      8\pi\,G\,e^{\lambda }\,P\,\left( x^2 + y^2 + z^2 \right)  \right) }{c^3\,\left( x^2 + y^2 + z^2 \right) }
\end{eqnarray}}
According to the definition (\ref{tensorM}) one finds the 00 component to be
\begin{equation}
M_0^0=\frac{c^4}{8\pi G}\left( \partial_x\chi_{0}^{01} + \partial_y\chi_{0}^{02} + \partial_z\chi_{0}^{03} \right)
\end{equation}
After taking partial derivatives of potential $\chi_{0}^{0\gamma}$ and using Eqs. (\ref{me1}, \ref{me2}, \ref{me3}) the M\o ller complex reads
{\setlength\arraycolsep{2pt}
\begin{eqnarray}
\nonumber
\lefteqn{M_0^0= \frac{e^{(\lambda  + \nu)/2}}{-4\,c^3\,G\,M(r) + 2\,c^5\,r}\left( -8\,c^2\,G\,P^2(r)\,\pi \,r^3 + c^8\,r\,\rho(r)\right. } \\
\nonumber
& + & e^{\lambda }\,\left( -2\,G\,M(r) + c^2\,r \right) \,\left( c^4 + 8\,G\,P(r)\,\pi \,r^2 \right) \,\left( P(r) + c^2\,\rho(r)  \right) \\
\nonumber
& + &   c^6\,\left( 5\,P(r)\,r - 4\,G\,M(r)\,\rho(r)  \right) \\
& - & \left. 4\,c^4\,G\,P(r)\,\left( 3\,M(r) + 2\,\pi \,r^3\,\rho(r)\right)  \right)
\end{eqnarray}}
Using the definition of the function $\lambda$ and rearranging terms we obtain
\begin{equation}
M_0^0=c\,e^{\nu/2}\,{\sqrt{\frac{c^2\,r}{-2\,G\,M(r) + c^2\,r}}}\,\left( 3\,P(r) + c^2\,\rho(r)  \right).
\end{equation}
The M\o ller pseudotensor is $M_0^0 = \sqrt{-g}(t_0^0 - T_0^0)$ and one can easily find that in this prescription the 
gravitational energy density is given by very simple formula $t_0^0=3\,P(r)$.

\section{Landau-Lifshitz prescription}
For the metric in Cartesian coordinates we have
\begin{equation}
L^{00}=\frac{c^4}{16\pi G}\partial_{\rho}\partial_{\sigma}\left(-g g^{00}g^{\sigma\rho}\right)
\end{equation}
The determinant of the metric tensor is $-e^{\nu+\lambda}$ and the contravariant component $g^{00}$ is $e^{-\nu}$, so
Landau-Lifshitz complex reads
{\setlength\arraycolsep{2pt}
\begin{eqnarray}
\nonumber
L^{00}=\frac{c^4}{16\pi G}\left(\partial_x^2 e^{\lambda}g^{11} + \partial_y^2 e^{\lambda}g^{22} + \partial_z^ 2e^{\lambda}g^{33} \right)\\
+\frac{c^4}{8\pi G}\left(\partial_x\partial_y e^{\lambda}g^{12} + \partial_x\partial_z e^{\lambda}g^{13} + \partial_y\partial_z e^{\lambda}g^{23} \right)
\end{eqnarray}}
Inserting the components of contravariant metric tensor, taking partial derivatives and using 
Eqs. (\ref{me1}, \ref{me2}, \ref{me3}) we obtain
\begin{equation}
L^{00} = \frac{c^4\,\left( -G\,M^2(r)  + 2\,c^2\,\pi \,r^4\,\rho(r)  \right) }{2\,\pi \,r^2\,{\left( -2\,G\,M(r) + c^2\,r \right) }^2}.
\end{equation}
For $r>R$ we have the Schwarzschild solution. So taking $\rho=0$ we find how does the Landau-Lifszyc complex look like in 
the vacuum outside the star: the vacuum energy density is negative. The same result was obtained by Virbhadra 
(see Eq.(16) in \cite{Virbhadra}  with $Q=0$ and wiht $G=c=1$). Because $L_0^{0} = L^{00}g_{00}$ we finally find the 
gravitaional field energy density in 
Landau-Lifshitz form to be
\begin{equation}
t_0^0=\frac{c^2\,G\,M(r)\,\left( M(r) - 4\,\pi \,r^3\,\rho(r)  \right) }{2\,\pi \,r^3\,\left( 2\,G\,M(r) - c^2\,r \right) }.
\end{equation}

\section{Weinberg pseudotensor}
In the Weinberg prescription \cite{Weinberg} the complex of matter and field energy density is
\begin{equation}
W^{\mu\nu}=\frac{c^4}{16\pi G}{\Delta^{\mu\nu\alpha}}_{,\alpha}
\end{equation}
The tensor $\Delta^{\mu\nu\alpha}$ is given by 
\begin{equation}
\Delta^{\mu\nu\alpha} = \frac{\partial h_{\beta}^{\beta} }{ \partial x_{\mu} } \eta^{\nu \alpha} -							\frac{\partial h_{\beta}^{\beta} }{ \partial x_{\nu} } \eta^{\mu \alpha} - \frac{\partial h^{\beta\mu} }{ \partial x^{\beta} } \eta^{\nu \alpha} +\frac{\partial h^{\beta\nu} }{ \partial x^{\beta} } \eta^{\mu \alpha} + 		\frac{h^{\mu\alpha}}{\partial x_{\nu}} - \frac{h^{\nu\alpha}}{\partial x_{\mu}}
\end{equation}
where $h_{\mu\nu}=g_{\mu\nu} - \eta_{\mu\nu}$, and $\eta_{\mu\nu}$ is Minkowski metric. We look for $W^{00}$. Using $h_{\beta}^{\beta}=\eta^{\mu}_{\beta}h_{\beta\nu}$ we find that
{\setlength\arraycolsep{2pt}
\begin{equation}
\Delta^{00\alpha} = \frac{\partial h_{\beta}^{\beta} }{ \partial x_0} \eta^{0 \alpha} - \frac{\partial h_{\beta}^{\beta} }{ \partial x_0 } \eta^{0 \alpha}- \frac{\partial h^{\beta0} }{ \partial x^{\beta} } \eta^{0 \alpha}\\
+\frac{\partial h^{\beta0} }{ \partial x^{\beta} } \eta^{0 \alpha} + \frac{\partial h^{0\alpha} }{\partial x_0} - \frac{\partial h^{0\alpha} }{\partial x_0}=0
\end{equation}}
So the Weinberg complex 00 component is equal 0 for any $\alpha$ and $\beta$ as it was observed by Xulu \cite{Xulu}.

\end{document}